\documentclass[epj]{svjour}
\usepackage{latexsym}
\usepackage{graphics}
\begin{document}
\title{The kaon optical potential modified by $\Theta^+$ Pentaquark excitation}
\author{D.~Cabrera\inst{1} \and L.~Tol\'os\inst{2} \and A.~Ramos\inst{3} \and
A.~Polls\inst{3}
}                     
%
%
\institute{\inst{1}Cyclotron Institute and Physics Department, Texas A\&M University,
College Station, Texas 77843-3366, U.S.A.\\
\inst{2}Gesellschaft f\"ur Schwerionenforschung,   
Planckstrasse 1, D-64291 Darmstadt, Germany\\
\inst{3}Departament d'Estructura i Constituents de la Mat\`eria,
Universitat de Barcelona, Diagonal 647, 08028 Barcelona, Spain
}
\date{Received: date / Revised version: date}
%
\abstract{
We study the 
kaon nuclear optical potential including the effect of the
$\Theta^+$ pentaquark in a selfconsistent approach, starting from
an extension of the J\"ulich meson-exchange
potential as bare kaon-nucleon interaction. 
Significant differences between a
fully self-consistent calculation and the low-density $T\rho$
approximation are observed. 
Whereas the influence of the $K N \to \Theta^+$ absorption process 
is found to be negligible, due to the small $\Theta^+ K N$ coupling, 
the two-nucleon mechanism ($K N N \to
\Theta^+ N$), estimated from the two-meson coupling of the pentaquark, provides
the necessary additional kaon absorption to reconcile with data the
systematically low theoretical $K^+-$nucleus reaction cross sections.
\PACS{
      {13.75.Jz}{}   \and
      {25.75.-q}{}   \and
      {21.30.Fe}{}   \and
      {21.65.+f}{}   \and
      {12.38.Lg}{}   \and
      {14.40.Ev}{}   \and
      {25.80.Nv}{}
     } 
} 
\maketitle
\section{Introduction}
\label{sec:intro}

The possible existence of the $\Theta^+$ pentaquark with positive strange\-ness
\cite{osaka,hyodo}  and a very
small width has attracted renovated interest on the properties of the $KN$
interaction \cite{haidenbauer,sibirstev}.
The medium properties of kaons are of
particular interest as they are considered suited probes of the hot and dense
matter created in heavy ion collisions
and of partial restoration of chiral symmetry in dense matter. The in-medium
$KN$ interaction, believed to be smooth in the absence of  baryonic resonances
with $S=+1$, has been usually implemented in a $T \rho$ approximation,
leading to a repulsive single-particle kaon potential of around 30~MeV at
normal nuclear matter density~\cite{kaiser,oset} (the more recent
self-consistent approach in \cite{korpa} showing a kaon mass shift of 36~MeV at
$\rho=\rho_0$). However, calculations of the kaon optical potential based on the 
$T\rho$ approximation share a common lack of success in reproducing
$K^+-$nucleus total and reaction cross sections (see
\cite{friedman} and references therein),
underestimating the data by about 10-15\%, a problem which remains unsolved
although a wide variety of mechanisms has been explored
\cite{siegel,brown2,jiang,garcia}.

In this work we start by comparing the $T \rho$ approximation of
the kaon optical potential to a fully self-consistent calculation of
the in-medium $KN$ effective interaction.  The latter, based on the J\"ulich
meson-exchange model including the coupling of the $\Theta^+$ to the $KN$ system
\cite{haidenbauer}, allows us to investigate the changes on the kaon optical
potential due to the one-nucleon induced excitation of the 
$\Theta^+$ in the medium.
In addition, following the suggestion of \cite{gal}, in which
substantially improved fits to the data were obtained by incorporating 
$K^+$ absorption by nucleon pairs ($K N N \to \Theta^+ N$), we evaluate  a
microscopic calculation of this mechanism according to the model of
Ref.~\cite{hosaka} for the $\Theta^+$ interaction with a $K \pi$ cloud, where
the 
pion couples to particle-hole ($ph$) and Delta-hole ($\Delta h$) excitations. We
study the effect of this mechanism in the kaon optical potential and obtain
improved $K^+$ nuclear cross sections in close agreement with the experimental
data~\cite{Tolos:2005jg}.

\section{In-medium $KN$ interaction}
\label{sec:KN}

\begin{figure}
\begin{center}
\resizebox{0.15\textwidth}{!}{%
  \includegraphics{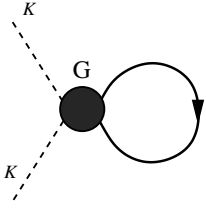}
}
\end{center}
\caption{One-nucleon contribution to the kaon self-energy from the in-medium
$G-$matrix.}
\label{fig:diagKN}
\end{figure}

We evaluate the in-medium $KN$ interaction in a $G-$matrix approach, including 
Pauli blocking on the nucleonic intermediate states and
the dressing of the $K$ meson and nucleon. Schematically,
\begin{eqnarray}
&&\langle K N \mid G \mid K N \rangle =
\langle K N
\mid V \mid K N \rangle   \nonumber \\
&+&  \langle K N \mid V \mid K N \rangle
\frac {Q_{K N}}{\Omega-E_{K} -E_{N}+i\eta} \langle K N \mid
G \mid K N \rangle \ .
\nonumber \\
   \label{eq:gmat1}
\end{eqnarray}
The bare interaction, $V$, is obtained from the meson-ex\-chan\-ge J\"ulich
model for $KN$ scattering with
a $\Theta^+$ pole term \cite{haidenbauer}. The $\Theta^+$ bare mass and
coupling to $KN$ are chosen as to reproduce the physical $\Theta^+$ mass and a width of 5~MeV
in free space (widths higher than a few MeV have been excluded by recent
analysis of the $K^+N$ and $K^+$d data \cite{diakonov,nussinov,arndt,gibbs}). 
The kaon single-particle energy in Eq.~(\ref{eq:gmat1}) is obtained
self-consistently as $ E_{K}(\vec{q};\rho)=\sqrt{m_{K}^2+\vec{q}\,^2} + {\rm Re}\,U_{K}
(E_{K},\vec{q};\rho)$,
with $U_{K}$ the complex single-particle potential which, in the
Brueckner-Hartree-Fock approach, is given by
\begin{eqnarray}
&& U_{K}(E_{K},\vec{q};\rho) = 
\nonumber \\
&& \sum_{N \leq F} \langle K
N \mid
 G_{K N\rightarrow
 K N} (\Omega = E_N+E_{K}) \mid  K N
\rangle \ , 
\label{eq:self0}
\end{eqnarray}
as diagrammatically represented in Fig.~\ref{fig:diagKN}.
The nucleon single-particle energies are taken from
a relativistic $\sigma-\omega$
model with density-dependent scalar and vector coupling
constants \cite{Mach89}, which provides an attraction of $-$80~MeV at saturation
density.

\begin{figure}
\begin{center}
\resizebox{0.4\textwidth}{!}{%
  \includegraphics{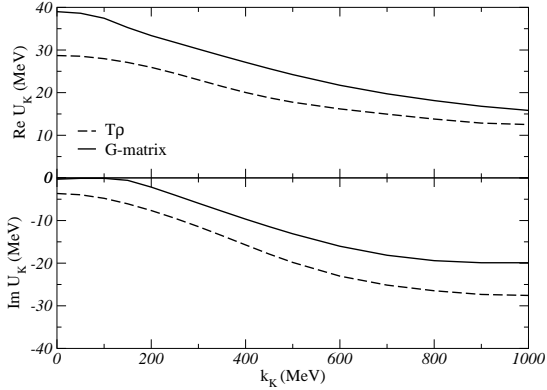}
}
\end{center}
\caption{Kaon optical potential as a function of
      the kaon momentum at normal nuclear matter density for the $T \rho$
      approximation (dashed lines) and the $G$-matrix calculation (solid
      lines).}
\label{fig:potential}
\end{figure}

We show in Fig. \ref{fig:potential} the kaon optical potential at $\rho=\rho_0$
in absence of the coupling to $\Theta^+$, for the self-consistent $G-$matrix
calculation and a $T\rho$ approximation (replacing $G$ in Eq. (\ref{eq:self0})
by the free amplitude $T$).
The latter, which ignores in-medium effects on the $KN$
interaction as well as on the $K$, $N$ energies, leads to a kaon optical
potential (at zero 
momentum) 10~MeV less repulsive than the $G-$matrix result, which amounts to
39~MeV. The imaginary part of the potential indicates kaon widths of about
18~MeV at about 400~MeV$/c$ momentum. The selfconsistent implementation of medium
effects generates relevant features of the $KN$ interaction which
reflect on the kaon properties in the nuclear medium in comparison with the 
$T\rho$ approximation.

%
%

%

\section{Kaon absorption from $\Theta^+$ excitation}
\label{sec:Kabsorption}

\begin{figure}
\begin{center}
\resizebox{0.33\textwidth}{!}{%
  \includegraphics{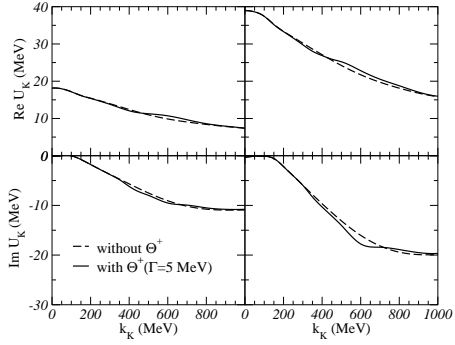}
}
\end{center}
\caption{Kaon optical
       potential for
       $0.5$ $\rho_0$ (left panels) and $\rho_0$ (right panels), without
       $\Theta^+$ (dashed lines) and
       including a $\Theta^+$ resonance with a width of 5
       MeV (solid lines).}
\label{fig:pentaone}
\end{figure}

Turning on the $KN\Theta^+$ coupling incorporates the effect of one-body
$\Theta^+$ excitation in the effective $KN$ interaction. Interestingly, the
$G-$matrix exhibits the expected structure associated to the $\Theta^+$ with
practically unchanged properties (mass and width) up to $\rho=\rho_0$ as
compared to the free 
$T-$matrix amplitude, except for the in-medium modified $KN$ threshold. This
result is highly non-trivial since the properties of the $\Theta^+$ in this
approach result from multiple rescattering in the $T-$matrix equation and
interferences between the polar and non-polar terms of the $KN$
interaction. The selfconsistent kaon optical potential (Fig.~\ref{fig:pentaone})
exhibits rather small changes (starting above 300~MeV$/c$ momentum) from the $KN\to\Theta^+$
mechanism, which is related to a small $KN\Theta^+$ coupling (as implied by the
small $\Theta^+$ width).

The kaon optical potential may also receive contributions from the two-nucleon
process $K NN \to \Theta^+ N$ which can be realized from the $\Theta^+$ coupling
to $K\pi$ with subsequent absorption of the pion
by a $ph$ or $\Delta h$ excitation (Fig.~\ref{fig:diagtwo}). The $\Theta^+
K\pi N$ coupling was evaluated in a study of the two-meson cloud effects on the
baryon antidecuplet binding in vacuum \cite{hosaka} and latter applied in a
calculation of the $\Theta^+$ self-energy in nuclear matter \cite{Cabrera:2004yg}. 
Its
contribution to the kaon optical potential can be obtained from the kaon
self-energy diagram in Fig.~\ref{fig:diagtwo} (right),
\begin{eqnarray}
\Pi^{2N}_K(q^0,\vec{q};\rho) = i \int\frac{d^4k}{(2\pi)^4}
\left[D_\pi^{(0)}(k)\right]^2
\Pi_\pi(k;\rho)
\nonumber \\
 \times (-9) \sum_{j=S,V}  \mid t^{(j)}(k,q) \mid^2
U_\Theta(q-k;\rho) \ ,
\label{eq:selfk2N}
\end{eqnarray}
where $D_\pi^{(0)}(k)$ is the free pion propagator,
$\Pi_\pi(k;\rho)$ stands for the $ph+\Delta h$ contribution to the
pion self-energy including short range correlations, $U_\Theta(q-k;\rho)$
represents
the pentaquark-hole Lindhard function and $t^{(j)}$ denotes the scalar ($S$)
and vector ($V$) $\Theta^+ K\pi N$ amplitudes \cite{Tolos:2005jg}.
The contribution of this mechanism to the imaginary part of
the kaon optical potential is shown in Fig.~\ref{fig:pentatwo} as a function of
the density for a kaon momentum of 500~MeV$/c$, exhibiting the expected
$\rho^2$ dependence of a two-nucleon process and a significant size, comparable 
to the one-nucleon mechanisms depicted in Fig.~\ref{fig:potential}.

\section{$K-$nucleus cross section}
\label{sec:cross}

\begin{figure}
\begin{center}
\resizebox{0.37\textwidth}{!}{%
  \includegraphics{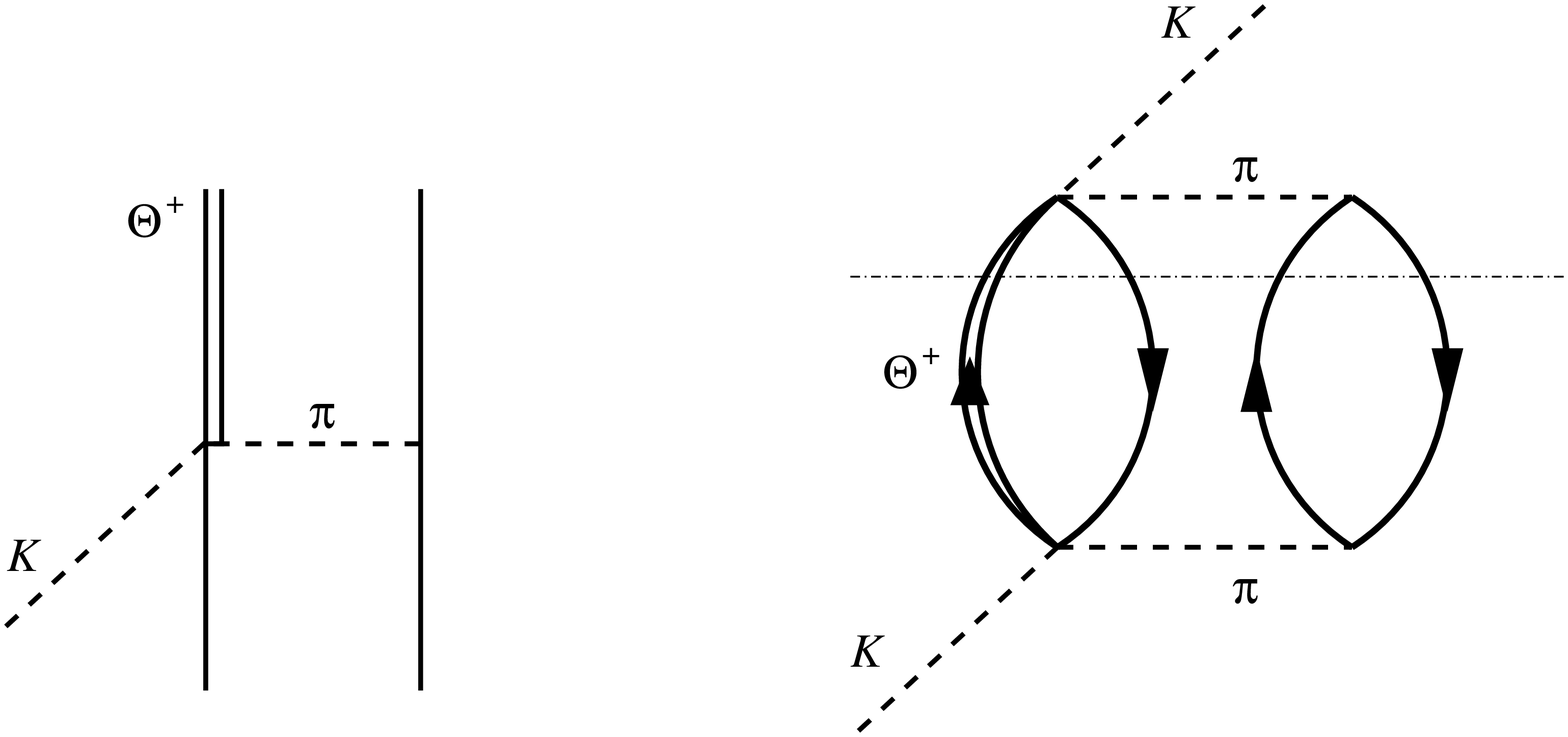}
}
\end{center}
\caption{Two-nucleon kaon absorption mechanism by excitation of $\Theta^+$
(left) and corresponding manybody diagram (right).}
\label{fig:diagtwo}
\end{figure}

\begin{figure}
\begin{center}
\resizebox{0.33\textwidth}{!}{%
  \includegraphics{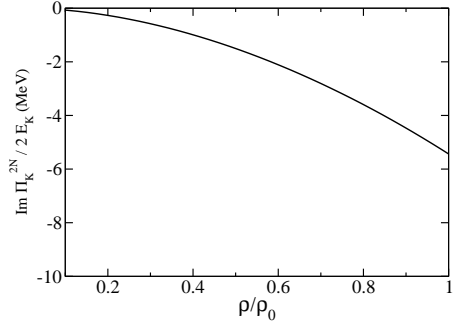}
}
\end{center}
\caption{Imaginary part of the two-nucleon contribution to the kaon optical
potential as a function of the density for a kaon momentum of 500~MeV$/c$.}
\label{fig:pentatwo}
\end{figure}

\begin{figure}
\begin{center}
\resizebox{0.33\textwidth}{!}{%
  \includegraphics{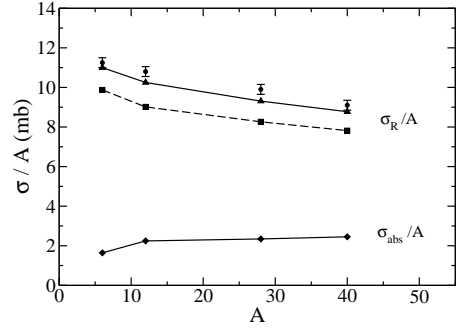}
}
\end{center}
\caption{Absorption and reaction cross sections per nucleon for a kaon
      laboratory
      momentum of 488 MeV/c, from the $G-$matrix kaon
      optical potential (dashed line) and including, in addition,
      the $2N-$absorption mechanism (solid line).
      Data for $\sigma_{\rm R}/A$ are taken from the
      analysis of \protect\cite{friedman}.}
\label{fig:cross}
\end{figure}

We calculate kaon nuclear cross sections in the eikonal formalism 
according to
\begin{equation} 
\sigma= \int d^2b \left[ 1 - {\rm exp}\left( -
\int_{-\infty}^\infty -\frac{1}{q} {\rm Im}\, \Pi(q;\rho(\vec{b},z)) dz \right)
\right] \ , 
\label{eq:cross1}
\end{equation} 
where $\Pi$ is given by the two-nucleon component of
the kaon self-energy ($\Pi^{2N}_K$) for the absorption cross section
($\sigma_{\rm abs}$), or by the total self-energy including
as well the contribution of the one-nucleon mechanisms for the reaction cross
section ($\sigma_{\rm R}$), respectively.
In Fig.~\ref{fig:cross} the calculated cross sections per nucleon
in $^6$Li, $^{12}$C, $^{28}$Si and $^{40}$Ca are compared to experimental data.
The results for $\sigma_{\rm R}/A$ from the $G-$matrix calculation of the kaon optical potential
including the one-nucleon $\Theta^+$ excitation mechanism underestimate the data
by about 15\% (dashed lines).
On the other hand, the absorption cross sections per nucleon
obtained from the $2N$ mechanism (Figs. \ref{fig:diagtwo} and 
\ref{fig:pentatwo}) are about 2-3 mb, right below the upper bound
of 3.5~mb established in Ref.~\cite{gal}.
The reaction cross sections per nucleon obtained with the
complete imaginary part of the kaon self-energy, including both the one- and
two-nucleon processes, lie very close to the experimental data (upper solid
line).
Our
model for the two-nucleon kaon absorption mechanism provides the required
strength to bring the reaction cross sections in agreement with experiment,
thereby giving a  possible answer to a  long-standing anomaly in the physics of
kaons in nuclei.

%
%

%
%

%
%

%

\section{Acknowledgments}
\label{sec:acknow}
We thank
J. Haidenbauer for providing us with the extended J\"ulich
code, and L. Roca and M.J. Vicente-Vacas for fruitful discussions. 
L.T. acknowledges support from the
Alexander von Humboldt Foundation and D.C. from
Ministerio de Educaci\'on y Ciencia (Spain). This work is partly supported
by DGICYT contract BFM2002-01868, the Generalitat de Catalunya
contract SGR2001-64, and the E.U. EURIDICE network contract
HPRN-CT-2002-00311. This research is part of the EU Integrated
Infrastructure Initiative Hadron Physics Project under contract
number RII3-CT-2004-506078.

%
%

\end{document}